\documentstyle[prl,aps,epsfig,multicol]{revtex}

\newcommand{\fS}{f_{\rm S}}

\begin{document}


\draft

\title{Adaptive walks on time-dependent fitness landscapes}

\author{Claus O. Wilke and Thomas Martinetz}

\address{Institut f\"ur Neuroinformatik, Ruhr-Universit\"at Bochum,
  D-44780 Bochum, Germany}

\date{Submitted: ; Printed: \today}

\maketitle

\begin{abstract}
The idea of adaptive walks on fitness landscapes as a means of
studying evolutionary processes on large time scales is extended to
fitness landscapes that are slowly changing over time. The influence
of ruggedness and of the
amount of static fitness contributions  are
investigated for model landscapes derived from Kauffman's $NK$ landscapes.
Depending on the amount of static fitness contributions in the
landscape, the evolutionary dynamics can be divided into a percolating and a
non-percolating phase. In the percolating phase, the walker performs a
random walk over the regions of the landscape with high fitness. 
\end{abstract}



\begin{multicols}{2}
Most work on Darwinian evolution so far has been concerned with
evolution in constant environments on the one hand (e.g.,
see~\cite{Kimura84,Ewens79,Gale90} for the field of population
genetics, or~\cite{Kauffman92} for adaptive walks), and coevolutionary
processes on the other hand (e.g., see~\cite{Freedman80,HallamLevin86}
for ecological models like Lotka-Volterra systems,
or~\cite{Hillis91,KauffmanJohnsen91,Reynolds94} for Artificial Life
type computer simulations). The case in which a species is subjected
to a changing environment, without being able to influence it, has
been studied only rarely. Most work of the latter type is considering
a single periodically changing
optimum~\cite{Charlesworth93,LandeShannon96,IshiiMatsudaIwasaSasaki89,Baeck98}.
In such situations, the evolutionary dynamics acts as a low pass
filter~\cite{HirstRowe99}. The optimum can only be tracked if the
oscillation frequency is sufficiently low.

In this work, we are considering evolution in high-dimensional
fluctuating fitness landscapes, with different amounts of dynamic and
static fitness contributions. The motivation of this work comes from
\emph{in vivo} evolution of proteins. Living organisms use a huge
amount of different proteins. Where does this diversity originate
from? When looking at a single protein in a particular species, the
protein appears to be in a local optimum, without any better mutants
nearby. However, to account for the observed diversity, there must be
mechanisms that allow to move on from one local optimum to another
sporadically. The simplest mechanism one can consider is one in which
large mutations sometimes carry a protein into a distant region in the
genotype space. Although this mechanism cannot completely be rejected, it is
unlikely that large mutations play a predominant role 
in protein evolution. A large mutation is essentially a random jump
into the genotype space, leading with extremely high probability to an
amino acid sequence that cannot fold correctly anymore. Therefore,
large mutations will in almost all cases not produce a viable
protein. 

A mechanism that works with small mutations is drift on
neutral networks. It has been mostly studied for 
RNA~\cite{ForstReidysWeber95,HuynenStadlerFontana96,ReidysStadlerSchuster97},
but it can also be considered in the case of
proteins~\cite{BabajideHofackerSipplStadler97}. On a neutral network,
mutations change 
the amino acid sequence, but leave the protein fold and, more
importantly, the protein's active region unaltered. From time to time,
the drifting sequence comes close to a sequence with higher fitness,
and then a transition to a new local optimum takes place. This theory
works well for \emph{in vitro} experiments~\cite{Forst98}, but it is
unclear whether enough neutrality exists \emph{in vivo} to allow for
sufficient drift~\cite{BennerEllington88,KreitmanAkashi95,Kreitman96}. There
exists evidence that in some cases, no neutral amino acid
substitutions exist in living organisms (e.g., for 
\emph{Drosophila m.}'s alcohol dehydrogenase locus~\cite{Kreitman83})
and that the environment can 
select for extremely small fitness
differences~\cite{BerryKreitman93,Thatcheretal98}. The reason why this
cannot be observed in \emph{in vitro} experiments is probably that the
experiments are not sensitive enough~\cite{Thatcheretal98}.

Benner and Ellington~\cite{BennerEllington88} have
suggested a different mechanism that could work with small mutations
and in the absence of neutrality. They propose that slow environmental
changes generate a constant genetic drift which can be accounted for
the protein diversity. This idea has never been studied quantitatively
in a mathematical model.

Here, we are going to study a model which demonstrates that indeed 
a slowly changing environment can generate something like a constant
genetic drift. We will call this drift ``environmentally guided
drift''. It is not a diffusion process, such as neutral
drift~\cite{Kimura84}. The population as a whole moves through the
genotype space, since transitions to selectively advantageous states
happen very fast, as first order phase
transitions~\cite{Gillespie83,Adami98}. Adaptive walks
are particularly suitable to study such phenomena, and we will use
them  throughout this paper, neglecting population effects or
crossover of genotypes. 

The statement that the population always remains located in the
genotype space, and that hence the dynamics can be approximated with an
adaptive walk, can only be justified if the environmental changes are very
slow. If this is the case, i.e., if the fitness landscape does change
only marginally over time intervals of the length of typical waiting
times between two transitions, the adaptive walk approximation should
be valid under the same assumptions as in static landscapes. Note that
this means, on the other hand, that in our model the adaptive walk
must be allowed to do a number of jumps prior to significant changes in
the landscape. As a consequence, the walker will
often have the chance to reach a local optimum  before it
starts out for a new peak because of the deforming landscape.
Later in this paper, we will discuss the adiabatic limit,
which is an even slower time scale. In the adiabatic limit, the
changes happen so slowly that for every change in the fitness
landscape the adaptive walk can always find a local optimum.

As example landscapes, we choose Kauffman's $NK$
landscapes~\cite{Kauffman92,KauffmanWeinberger89}, which are  
spin glass-like 
landscapes commonly used for the  study of  adaptive walks.
Although they cannot be directly related to the true landscapes
underlying \emph{in vivo} protein evolution, their tunable degree of
ruggedness makes them a good tool to study general effects in rugged
landscapes. In a $NK$ landscape, the fitness of a bit string of length
$N$ is defined as 
the average over each bit's fitness contribution.
The contributions depend on the state of the corresponding bit as well
as on the state of $K$ other bits interacting with it. We
can write the fitness $f$ as:
\begin{equation}\label{eq:fitness-NK}
  f = \frac{1}{N} \sum_{i=1}^{N} f_i(\{S\}_i)\,,
\end{equation}
where $\{S\}_i$ is the state of the $K+1$ bits influencing the fitness
contribution of bit $i$. In Kauffman's
original formulation, the functions $f_i(\{S\}_i)$ are realized as
tables containing a different quenched random fitness contribution for every
state $\{S\}_i$.
Here we are interested in a time-dependent form of 
Eq.~(\ref{eq:fitness-NK}). One possibility has been proposed by
Levitan and Kauffman~\cite{LevitanKauffman95}, who
have studied the case that the fitness cannot be measured exactly. In
their work, the measured fitness $f'$ equals the true fitness $f$ plus
a noise term $g(t)$. Such an approach has proven suitable to study the
effects of noise in chemical engineering
problems~\cite{Levitan97}. However, it does not work here, since we are
interested in local deformations of the landscape, and not in a global
noise-induced shift. What we do instead is to use time-dependent functions
$f_i(\{S\}_i, t)$. Note the general difference in the
model of Levitan and Kauffman and of ours. In their model, the
fitness landscape is static, but can only be measured with finite
accuracy. In our model, the fitness landscape itself is
changing, but the fitness can be measured exactly. Considering the
long time scale we are addressing,  we 
can assume that noise does not play a prominent role here. The
single walker represents the mean of a population, as noted above. In
the population mean,  the noise is averaged out. We
will later discuss how the model could be altered for noise too intense
to allow that assumption, or for populations so small that
destabilizing effects can occur.

We choose the functions $f_i(\{S\}_i, t)$ to be
continuous in time. Noisy, discontinuous fitness contributions seem to
be unappropriate to model a slowly changing environment. In principle,
one could of course add a noise term on top of each fitness
contribution, or study landscapes with mixed noisy and continuous
contributions, but this is not our objective here.

Not necessarily all fitness contributions need to be truly time
dependent. Some may be equal to a constant,
\begin{equation}
  f_i(\{S\}_i, t)=C{_i,\{S\}_i}\,.
\end{equation}
It is useful to keep track of the amount of static contributions in
the landscape. We denote the fraction of static contributions by
$\fS$. Adaptive walks on time-dependent $NK$ landscapes show
several distinct classes of behavior, most strongly influenced by
$\fS$.

So far we have described the basis of our model, now we have to
specify the actual form of the fitness contributions. For
data analysis, it is useful to impose periodic boundary
conditions on the fitness landscape, i.e.,
\begin{equation}
  f(t+T) = f(t)\,,
\end{equation}
with oscillation period $T$. Throughout the rest of this work, we will
stick to this choice. Its advantage rests in the easy comparison of a
bit string's evolution in different oscillation periods. In particular,
it allows to use the concept of environmentally linked networks
introduced below.

The form of the functions $f_i(\{S\}_i,t)$ can in principle be
arbitrarily complicated. We are going to consider a simple trigonometric time
dependency, 
\begin{equation}
f_i(\{S\}_i,t)=\frac{1}{2}[\sin(\omega_{\{S\}_i} t +
\delta_{\{S\}_i})+1]\,.
\end{equation}
This introduces only a single additional 
constant per fitness contribution, if compared to the static
landscape. The frequencies $\omega_{\{S\}_i}$ and 
the phases $\delta_{\{S\}_i}$ are chosen randomly when constructing
the landscape, and are then kept fixed for all times $t$. The 
phases are distributed uniformly in the interval 
$[0; 2\pi)$ so that the resulting fitness landscape is 
homogeneous in time. We set the frequencies
to $\omega_{\{S\}_i}=2\pi n_{\{S\}_i}/T$, with
$n_{\{S\}_i}$ being integral, and $T$ being arbitrary, but the same for all
$\omega_{\{S\}_i}$, to 
obtain a periodic fitness landscape with oscillation period
$T$. If we want a fitness contribution to be constant, we set the
corresponding frequency $\omega_{\{S\}_i}$ to 0.

\begin{figure}
\narrowtext
\centerline{ 
 \epsfig{file={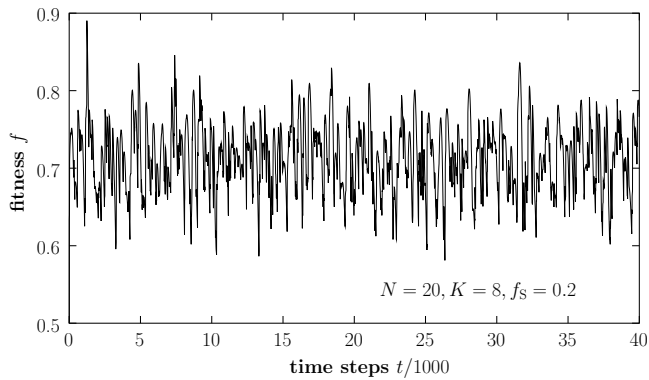}, width=\columnwidth}
}
\caption{The evolutionary dynamics is chaotic for small $\fS$.
\label{fig:chaos}}
\end{figure}

\begin{figure}
\narrowtext
\centerline{
 \epsfig{file={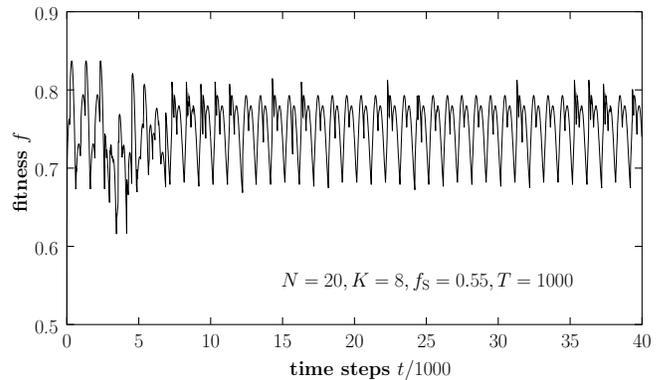}, width=\columnwidth}
}
\caption{With increasing $\fS$, some bits in the bit string
  freeze out, and the evolutionary pattern becomes more and more oscillatory.
\label{fig:oscill}}
\end{figure}

\begin{figure}
\narrowtext
\centerline{
 \epsfig{file={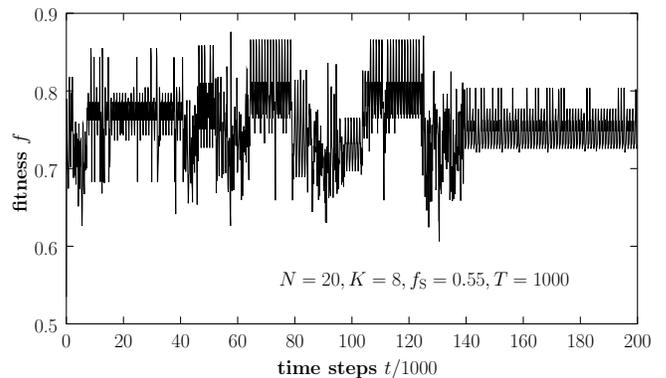}, width=\columnwidth}
}
\caption{The oscillatory states are metastable, and transitions
  between them can occur.
\label{fig:metastable}}
\end{figure}

We have done a large number of simulations with different choices for
$N$ and $K$, with different sets of oscillation frequencies, and also
with more complicated functions $f_i(\{S\}_i,t)$, in which the
oscillations have additional random amplitudes and offsets. In all
cases, the basic patterns are very similar. The parameters having the
strongest influence on the observed behavior are the ruggedness $K$
and the fraction of static fitness contributions $\fS$. In
Figs.~\ref{fig:chaos}-\ref{fig:metastable},  some typical 
runs of adaptive walks in oscillating $NK$-landscapes are
presented. In the simulations leading to these plots, we used $N=20$ and
$K=8$. Additionally, we employed only a single oscillation mode. This
means, all frequencies $\omega_{\{S\}_i}$ were either set to zero or
set to some fixed value $\omega=2\pi/T$. The oscillation period $T$
was set to $T=1000$, which can be considered large in a system with
$N=20$. A local optimum can typically be found in about 100 time steps
in a static $NK$ landscape with such $N$.
The adaptive walk 
was performed exactly as in Kauffman's original work: a random point mutation
was accepted if it increased the bit string's fitness. Otherwise, the
mutation was rejected.  

Figure~\ref{fig:chaos} shows an example of the 
evolutionary dynamics with a
relatively low fraction of static fitness contributions. The resulting
pattern is a chaotically changing fitness. With almost every accepted
mutation, 
a new genotype is encountered. The environmental changes
constantly lead the walker into regions previously not
visited. This reminds one of a random walk. However, there are some
differences between the adaptive walk and a random walk. We will
discuss them below.

The
behavior of the adaptive walk changes drastically with increasing
$\fS$. The higher amount of static fitness contributions reduces
the number of 
possible advantageous mutations in every time step. The bits connected
to static contributions freeze out in a locally optimal state, and
only the sites connected to oscillating contributions can still
change. Hence, the dynamics gets confined in a small region of the
genotype space. The same mutational patterns are seen over  and over again
in the different oscillation periods. In the fitness plots, we can
identify this behavior with a periodic or almost periodic change of
the fitness, as shown in Fig.~\ref{fig:oscill}. Using the language of
dynamic systems, we can say that the attractor of an adaptive walk
on an oscillating landscape with intermediate $\fS$ is a noisy limit
cycle. With some small probability $p$, the process can
leave a limit cycle. Several transitions between such metastable limit
cycles are shown in Fig.~\ref{fig:metastable}. The mean fitness  can
increase or decrease because of the transitions. The frequency with which
transitions occur depends on the actual value of $\fS$. The larger
$\fS$, the rarer can transitions be observed. 

These metastable states remind one very much of the metastability induced by
finite populations on static landscapes with a high degree of
neutrality~\cite{Nimwegenetal97,Nimwegenetal97a,CrutchfieldNimwegen99},
however they are generated through a completely different
mechanisms. A slight qualitative difference between the two types of
metastability is that here, the transitions lead regularly to a
decrease of a metastable state's average fitness, whereas in
neutrality-induced transitions, this is mostly not the case. Nevertheless,
the work of Nimwegen  \emph{et al.} shows that with very small 
populations, the evolutionary dynamics on a landscape with neutrality
can as well display transitions leading to a decrease of
fitness~\cite{Nimwegenetal97a}. The interesting point of our findings
here is that we find metastability under the
complete absence of neutrality. 

\begin{figure}
\narrowtext
\centerline{
 \epsfig{file={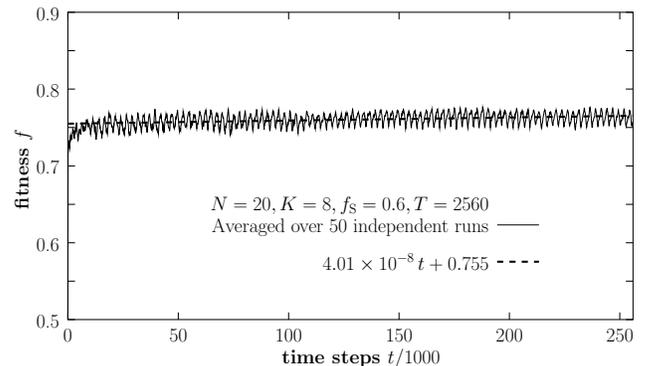}, width=\columnwidth}
}
\caption{Average fitness over time in adaptive walks over a landscape
  with $\fS=0.6$. The dashed line indicates the result of a least
  squares fit. The exact slope is
  $m=4.00824\times 10^{-8}\pm 2.057\times 10^{-10}$.
\label{fig:fitness-increase}}
\end{figure}

Let us now address the question whether the transitions actually lead
to an increase in fitness, or whether advantageous and disadvantageous
transitions balance each other. In Fig.~\ref{fig:fitness-increase}, we
show the expected fitness as a function of time for 100 oscillations
with a period of $T=2560$. The expected fitness was obtained by
averaging over 50 independent runs. We have chosen $f_{\rm S}=0.6$,
which is well in the metastable regime. We observe that the most
important fitness gain is reached during the first couple of
oscillations (the curve starts of from $\langle f\rangle = 0.5$ for
$t=0$). Nonetheless, for the complete duration of the 100 
oscillations, we observe a constant slight increase in the fitness. A
linear fit to the expected fitness from time step $10^4$ to the end of
the recording gives an increase in fitness of $1.03\times10^{-4}$ per
oscillation period. Ultimately, for much longer simulation runs, the
expected fitness reaches an asymptotic value. Note that the slight
fitness increase over many oscillation periods is an effect peculiar
to the metastable regime. In the chaotic regime the expected fitness
reaches its asymptotic value very quickly, after a few oscillation periods.

\begin{figure}
\narrowtext
\centerline{
 \epsfig{file={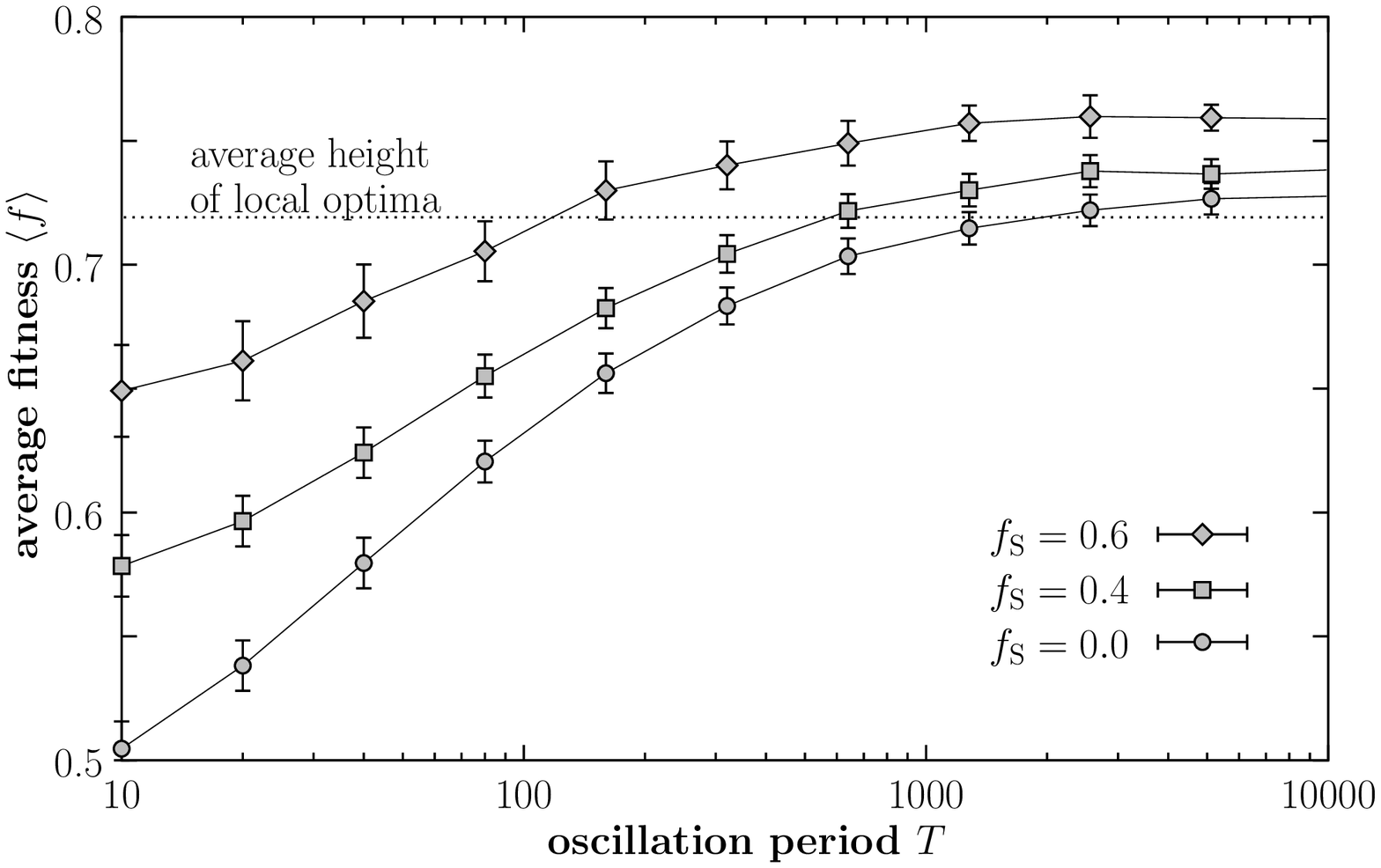}, width=\columnwidth}
}
\caption{Mean fitness encountered during an adaptive walk as a
  function of the oscillation period $T$. The fitness was averaged
  over 50 independent adaptive walks, of which each endured 100
  oscillation periods.
\label{fig:mean-fitness}}
\end{figure}

The adaptive walk's efficiency to find regions of high fitness can be
judged from the mean fitness encountered during the
walk. Figure~\ref{fig:mean-fitness} shows the mean fitness, averaged
over several independent adaptive walks, as a function of the oscillation
period $T$. The curve corresponding to the chaotic
regime, with $\fS=0$, starts off at a mean fitness of 0.5 for small
$T$. This is the average fitness on the landscape, and hence the
walker approximately does a random walk on the landscape. For larger
$T$, the mean fitness quickly grows and reaches a 
value close to the average of a local optimum in the
landscape. Although the movement in the genotype space appears to be
chaotic, the expected fitness of the walker at any point in time is as
large as the expected highest fitness an adaptive walk in a comparable
static landscape would encounter. Therefore, for large $T$ 
the walker's movement can be considered as a random walk confined to
the regions of high fitness in the genotype space. When we increase
the amount of static contributions in the landscape, the average
fitness is above 0.5 even for very fast environmental
changes. For larger $T$, the average fitness increases towards the
average height of local optima in the landscape, and even
significantly above it. The latter occurs
in time-dependent landscapes as long as only a
tiny amount of time-dependent contributions is present. To understand
why this happens, consider a
bit string in which all but one bit have only static contributions. The
remaining bit may also give a static contribution if it is set to 0, and
a time-dependent one if it is set to 1. For the times when the
time-dependent contribution is smaller than the static one, the bit will
be set to 0, and otherwise it will be set to 1. This effectively
increases the average height of local optima in dynamic
landscapes. The effect is most pronounced if the number of static
contributions is moderately large, for $\fS$ around 0.8.

At this point, it is interesting to ask what proportion of the
genotype space can actually be reached through environmentally guided drift.
The
question can be addressed with the concept of \emph{environmentally linked
networks} (EL networks). We define an EL network to be the set of all
points in the genotype space the adaptive walk can reach at times
$nT$, $n=0, 1, 2, \dots$, starting from a fixed position in the
genotype space. We will say an EL network percolates
if it consists of infinitely many points. This definition is similar
to the usual definition of the percolating cluster on the Bethe
lattice, and is the appropriate 
way to define percolation in high-dimensional
spaces~\cite{StaufferAharony92}. It can be applied literally
only in the limit $N\rightarrow\infty$. However, the genotype space
grows so fast with increasing $N$ that this restriction can be
neglected. 

If the walker is for small $\fS$ indeed doing a random walk over the
landscape, or over the landscape's regions of high fitness, as we
supposed above, then we should find a percolating EL network in the
chaotic regime.

The study of EL networks in oscillating $NK$ landscapes is
computationally very demanding, since we have to go through the full
oscillation periods in the simulation. Hence, we have to restrict
ourselves to moderate $T$ and $N$. In the examples below, we have
again used $\omega_i=\omega=2\pi/T$
with $T=1000$, as well as $N=20$.

\begin{figure}
\centerline{
 \epsfig{file={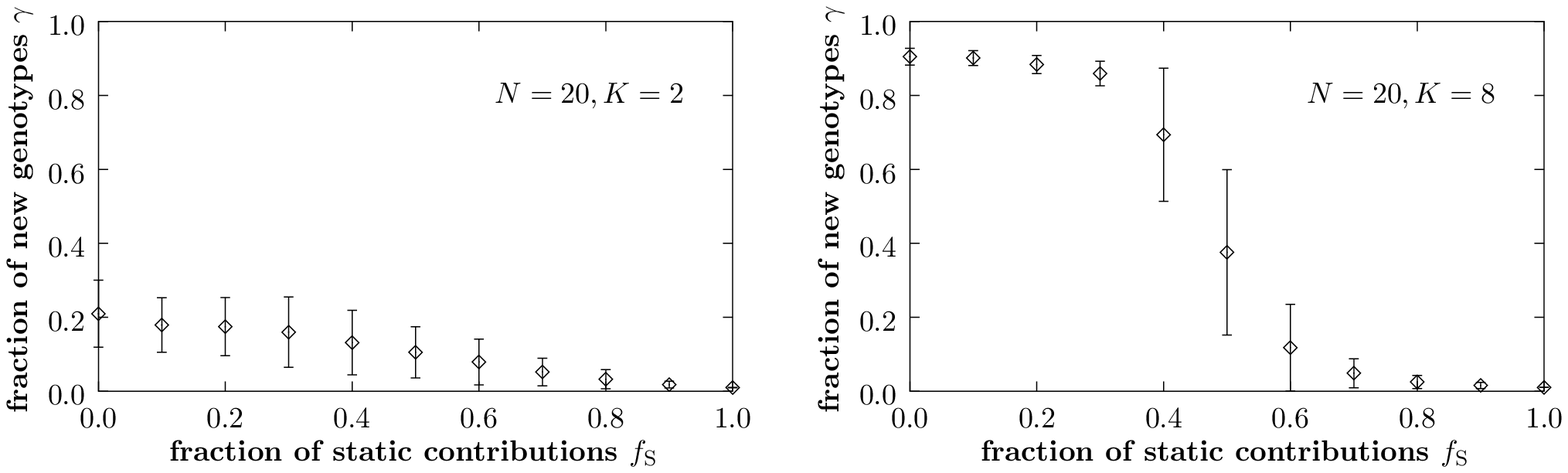}, width=\columnwidth}
}

\caption{Fraction of newly encountered genotypes $\gamma$ at the
  beginning of each oscillation period in oscillating fitness
  landscapes. The quantity $\gamma$ was averaged over 60 independent
  adaptive walks, of which each endured 200 oscillations with period
  $T=1000$.}
\label{fig:diversity}
\end{figure}

Figure~\ref{fig:diversity} shows the fraction $\gamma$ of new genotypes
among all the genotypes encountered at the beginning of each
oscillation period. This is a measure for the size of an EL network. A
value near 1 means a new genotype has been encountered in almost every
oscillation period. On the other hand, a value near 0 means the
network's size is small, thus confining the adaptive walk in a limited
region of the genotype space. In the limit of infinitely many
oscillation periods, only percolating networks can have a positive
$\gamma$, whereas finite networks yield $\gamma=0$. Therefore,
$\gamma$ is a proper order parameter indicating a percolation
transition. Clearly, in numerical experiments the number of
oscillation periods over which the measurement is taken is finite, and
therefore we will observe a positive $\gamma$ even in the
non-percolating regime. In the case of Fig.~\ref{fig:diversity}, the
value $\gamma$ was obtained from averaging over 60 
adaptive walks, each on a different fitness landscape. Every adaptive
walk endured 200 oscillation periods. The error bars present the
standard deviations of the single measurements.

Let us begin the discussion of Fig.~\ref{fig:diversity} with the
graph on the right, for $K=8$. We find a 
$\gamma$ close to 1 for small $\fS$, and a vanishing $\gamma$ for
$\fS\approx 1$. The standard deviations are very small in both limiting
regimes. In the region about $\fS\approx 0.5$, a sharp drop in
$\gamma$ can be observed, accompanied with an enormous increase in the
error bars. This is good evidence for the existence of a percolation
transition with critical point $\fS^\ast$ around 0.5. The large error
bars are a sign for critical fluctuations, 
observed in 2nd order phase transitions. The graph on the left
of Fig.~\ref{fig:diversity}, for $K=2$, shows a very different behavior.
The quantity $\gamma$ does not reach higher than about 0.2, and the
error bars are large for the whole range of $\fS$. We do not see a
clear percolation transition for this much less rugged landscape. The
large error bars indicate that the finite $\gamma$ for small $\fS$ is
rather an artifact due to the finite sampling than a true result. We
have done comparable simulations for the range of $K$ from 0 up
to 14, and what we generally observe is that the transition becomes
sharper with increasing $K$. 

We can understand the above observation in the
adiabatic limit.  For the case of a Fujiyama landscape ($K=0$), the EL
network degenerates to a single point in this limit, and percolation
can consequently not be observed. On the other hand, the completely
random landscape we obtain for $K=N-1$ presents a multitude of local
optima, and the changes in the landscape provide the opportunity to
hop from one local optimum to another in a random fashion during the
oscillation periods. The landscapes with intermediate $K$ interpolate
between the two extremes. This argument shows that ruggedness must
generally 
promote the movement in the genotype space for the low-$\fS$ regime, a
situation completely opposite to the case of static landscapes, where
ruggedness is regarded as an impediment to the movement in the
genotype space. If the changes happen on a slow enough time scale, the
increased mobility does \emph{not} lead to an error catastrophe
through which all information is lost, such as the breakdown of the
quasispecies for large mutation rates~\cite{Eigenetal89}. As we
saw in Fig.~\ref{fig:mean-fitness}, the fitness is constantly in the
region of average local optima. We observe this also in the example
run displayed in Fig.~\ref{fig:chaos}. The fitness is chaotically
changing, but it is always well above the landscape mean of 0.5. An
error catastrophe occurs only if the environmental changes happen very
fast compared to the typical adaptation time of the system.

So far, we disregarded noise or loss of information because of very
small populations in our model. Both have the effect to enable the
acceptance of mutations leading to lower fitness. Here, we could
incorporate them by adding noise to the fitness as
in~\cite{LevitanKauffman95}, or by accepting bad mutations with some
probability. As long as these mechanisms do not lead to an error
catastrophe on a static landscape, they should not very much alter the
dynamics on a slowly changing landscape from what we have found here.

Evolution in a slowly changing environment follows a dynamics very
different from the situation in a fixed environment. The environmentally
guided drift drives genes out of local optima, and drags them around
in the genotype space.  We have presented evidence for
the existence of a giant EL network for $\fS$ below some
critical $\fS^*$ in landscapes with sufficient
ruggedness. Consequently, in this regime the whole genotype space 
can be transversed by means of environmentally guided drift.
The guided drift can provide -- in absence of any
neutral pathways in the fitness landscape -- an efficient mechanism to
generate constantly new genotypes, albeit at every single
point in time the system seems to be trapped in a local optimum.
We could show that the efficiency of the
environmentally guided drift is related to the ruggedness of the
landscape. A more rugged landscape provides more opportunities to move
around under environmental changes than a landscape with only a
few peaks. Consequently, the rugged landscapes observed in protein
evolution~\cite{BennerEllington88} can promote protein evolution
in an ever changing environment, instead of hindering it. 
If we have a population that decomposes into several subpopulations
not coupled with each other through selection, these subpopulations
will disperse and move to completely different regions of the genotype
space because of environmentally guided drift, even if the process
starts off from a completely homogeneous population and if all
individuals in the system feel the same environmental changes at the
same time. The decoupling of the subpopulations can occur, for
example, if the population lives in a very large geographical
territory, so that individuals living in one part do not directly
interact with individuals living in another part, or if a physical
barrier forms at one point in time that divides the territory into
several independent regions. As a consequence, rugged landscapes
combined with slow environmental changes should inevitably lead to a
large variety of different evolutionary solutions for the same problems.

Although the EL networks used here for data analysis are only
meaningful in periodic landscapes, the conclusions drawn from their
study should also hold in non-periodic situations. The reason why
environmentally guided drift becomes so efficient for small
$\fS$ is that constantly new local optima appear nearby. Therfore, if
the changes are non-periodic, but the landscape has sufficient
ruggedness, the adaptive walk should similarly behave like a random walk
over the landscape's regions of high fitness.

An effect tightly connected to the periodicity of the landscape, on
the other hand, is the
appearance of limit cycles.  The dynamics in oscillating $NK$
landscapes is above the percolation transition dominated by noisy
limit cycles, with sporadically occuring transitions from one limit 
cycle to another. The system goes through several noisy limit cycles
until a stable limit cycle, or a stable fixed point, is reached. This
effect reminds one of the behavior of evolution on landscapes with a
high degree of neutrality. There, evolution proceeds on neutral
networks, with sporadic transitions between them, until a stable local
optimum is reached.

The model studied in the present paper, i.e., an adaptive walk on an
oscillating $NK$-landscape, is certainly too simplistic 
to be accounted for as a realistic model of the
\emph{in vivo} evolution of proteins in a changing environment. In
particular, it 
can be argued whether sinusoidally changing fitness contributions are
justified. Nevertheless, such simple models often capture the
qualitative properties of more realistic situations. Similar
percolation transitions can probably be found also in other
time-dependent landscapes with sufficient ruggedness.

In future work, it should be interesting to study the percolation
transition in more detail, and to determine for what $K$ a percolating
regime actually exists. Furthermore,  the interplay between  static
and dynamic fitness contributions should also be investigated in other
fitness landscapes.

\begin{acknowledgments}
We would like to thank S. Benner for introducing us to this field, T. Hirst for providing us with the manuscript of
Ref.~\cite{HirstRowe99}, C. Ronnewinkel for carefully
reading this manuscript, and T. Schmauch for double-checking the behavior
of oscillating $NK$ models for a number of different parameter settings.
\end{acknowledgments}


\end{multicols}

\widetext

\end{document}